\documentclass{winnower}
\usepackage{array}
\usepackage{multirow}
\usepackage{tabularx}
\usepackage[utf8]{inputenc}
\usepackage{natbib}
\usepackage{authblk}

\providecommand{\keywords}[1]
{
  \small	
  \textbf{\textit{Keywords- }} #1
}

\begin{document}

\title{\textbf{Study of involution domain based interfaces in Ni-Ti-Cu Shape Memory Alloy} } 

\author[1]{Atharva Pagare} 
\author[2]{V Sashi Mohan Rao} 
\author[3]{K Naga Chaithanya Kumar}
\author[4]{K S Suresh}
\affil[1,2,3,4]{\textit{Department of Metallurgical and Materials Engineering},\linebreak 
\textit{Indian Institute of Technology, Roorkee, India}}

\date{30th November, 2020}

\maketitle

\begin{abstract}
An algorithm was made to study the lattice correspondence involved in phase transformation from cubic B2 to monoclinic B19’. The method is based on studying the orientation matrices generated from EBSD data. Starting from the very fundamental, coordinate transformation matrices as well as the vector transformation matrices have been walked through for a general non-orthogonal to the orthogonal system. Further, using the defined formulas, orientation matrices will be used to identify a new, non-generic Involution Domain and the already accepted Bain Domain.

\end{abstract}


\section{Introduction}

Although a raft of research has been done on Ni-Ti alloys, it still struggles to reach the epitome of its efficiency that it is thought it contains. Motivated by the fact, a theoretical study has been done to gain insights into the kind of microstructure and phase transformation that occurs in nickel-titanium during mechanical deformation. Owing to a plethora of applications from biomedical to aerospace Ni-Ti has been of great importance to researchers. A constant struggle has always been going on to improve the mechanical properties of this alloy. Over decades scientists have tried to figure out the transformation mechanism from cubic austenite to monoclinic martensite phase. A theory created independently by Wechsler, Lieberman and Read(WLR Theory) \cite{1,2} and, Bowles and Mackenzie Theory \citep{3,4,5} is a very precise theory that predicts an essential parameter in the phase transformation, i.e. the habit plane. This habit plane is considered as an invariant plane upon the transformation that remains unrotated and undistorted. 

Nickel-Titanium amongst its wide range of applications exploit one of its significant properties of reversible solid to solid phase transformation. The reversibility depends significantly on the compatibility conditions between the phases, as shown in some studies \citep{6,7}. A very well accepted mechanism for transformation from a Pm3m symmetry cubic austenite phase to a P21/m symmetry monoclinic martensite phase is by a Bain-like transformation. The transformation is considered to have the least strain involved and the lattice correspondence i.e.

\begin{center}
$[1 0 0]_{B19'}\rightarrow[1 0 0]_{B2}\hspace{10pt}[0 1 0]_{B19'}\rightarrow[0 \bar{1} \bar{1}]_{B2}\hspace{10pt}[0 0 1]_{B19'}\rightarrow[0 1 \bar{1}]_{B2}$
\end{center}

is said as the Bain correspondence \citep{8,9}. The basal shear occurs on the [1 1 0] plane in the [0 0 1]  direction of the austenitic phase. Nothing that scientists already know about. However, interestingly a new correspondence was proposed recently with a transformation strain lower or much closer to the Bain-correspondence depending on the lattice parameter values \citep{10}. A sophisticated algorithm was developed to calculate different sub-lattices in the cubic phase that would transform into the conventional monoclinic lattice with a minimum transformation strain involved \citep{11}. StrucTrans algorithm defines a Distance Function dependent on the positive symmetric transformation stretch tensor U or called as Bain Matrix by Kaushik Bhattacharya in \citep{12}. The Distance function is made sure to converge to a finite value giving some non-standard results. A new kind of correspondence called Involution Domains is observed having transformation strain close to the generic Bain-correspondence. Depending on the lattice parameters, the transformation strain can be less or more than the Bain-correspondence. The detection of Involution Domains through X-Ray diffraction technique is incredibly difficult due to their exceptionally close lattice parameters to the Bain Domains. In this article, we try to observe the difference between the Bain and Involution Domains based on the orientation data of the Austenite and Martensite Phases.

There are multiple mathematical methods to describe the orientation of different phases or grains. Some of the major ones are by Miller Indices, Orientation Matrix, Euler Angle, Rodrigues Vector, Angle/Axis of Rotation. In order to describe an orientation, a reference coordinate system needs to be defined. The reference coordinate system is called as the sample/specimen reference frame with three linearly independent orthonormal bases which in material science we call as Rolling Direction (RD/x-axis), Transverse Direction (TD/y-axis), and Normal Direction (ND/z-axis). The first row of the orientation matrix O is defined as the cosines of the angle between the first crystal axis and the three sample axes {RD, TD, ND}, so on and so forth. Alternatively, we can also write the orientation matrix with the help of Miller Indices (h k l)[u v w] for a general crystal system. Orientation matrix is an orthonormal matrix, and writing it for any general crystal system involves some extra steps. For a crystal system with non-orthogonal bases there needs to be an orthonormal/cartesian basis $\{x^\#_c, y^\#_c, z^\#_c\}$ linked via a structure tensor A \citep{14, 15}.

Structure Tensor A can be defined arbitrarily and may change for different software manufacturers \citep{15}. In our study of EBSD data for a monoclinic crystal, the convention defined is as follows:

\begin{itemize}
    \item The z-axis is kept parallel to the c-axis of monoclinic.
    
    \item The y-axis is kept parallel to the b-axis of monoclinic.
    
    \item The x-axis is kept orthogonal to y and z axes.
    
    \item For a monoclinic system the angles are $\alpha = \gamma = 90^{\circ}, \beta > 90^{\circ}$.
    
    \item $\{a, b, c\}$ should be chosen according to the     right-hand rule.
\end{itemize}

Accordingly the Fig.~\ref{fig:f4}, used in the appendix, simplifies to Fig.~\ref{fig:f1}

\begin{figure}[h]
\begin{center}
    \includegraphics[scale = 0.4]{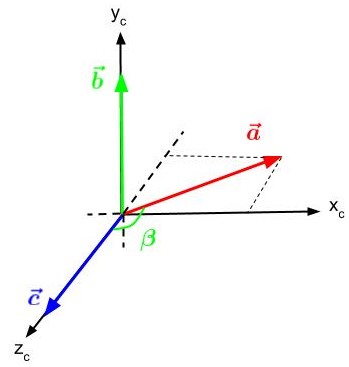}
    \centering
    \caption{Orthogonal reference frame(black) attached to a crystal reference frame(coloured).}
    \label{fig:f1}
\end{center}    
\end{figure}
\newpage

From the Fig.~\ref{fig:f1} Structure Tensor A is derived by writing the crystal bases {a, b, c} in the orthogonal reference frame $\{x^\#_c, y^\#_c, z^\#_c\}$ as the columns of the matrix:

\begin{equation}
    A = \begin{pmatrix}
    a\sin\beta   & 0 & 0\\ 
    0 & b & 0\\ 
    a\cos\beta  & 0 & c
    \end{pmatrix}    
\end{equation}

A normalised crystal plane normal (h k l) parallel to the samples ND direction and a crystal direction [u v w] parallel to the samples RD direction is not directly used to define an Orientation matrix. The crystal plane and directions are written in an orthogonal/cartesian frame first via a coordinate transformation. Thereafter, the plane and direction are written as the 3rd and the 1st column of the matrix, respectively.

Any direct lattice column vector $[u\,v\,w]_c$ in the crystal frame can be transformed into a vector
$[u\,v\,w]_o/[u_o v_o w_o]$ in an orthogonal frame (defined arbitrarily) by:

\begin{equation}
    \begin{pmatrix}
    u\\ 
    v\\ 
    w
    \end{pmatrix}_{o} = A\begin{pmatrix}
    u\\ 
    v\\ 
    w
    \end{pmatrix}_{c}  
\end{equation}

Similarly, any reciprocal lattice column vector $[h\,k\,l]_c$ in the crystal frame can be transformed into a vector $[h\,k\, l]_o/[h_o k_o l_o]$ in an orthogonal frame by:

\begin{equation}
    \begin{pmatrix}
    h\\ 
    k\\ 
    l
    \end{pmatrix}_{o} = A^{-T}\begin{pmatrix}
    h\\ 
    k\\ 
    l
    \end{pmatrix}_{c}  
\end{equation}                                              

Now, the orientation matrix O for a general crystal can be written as:
\begin{equation}
    O=\begin{pmatrix}
    u_{o} & d_{o} & h_{o}\\ 
    v_{o} & f_{o} & k_{o}\\ 
    w_{o} & g_{o} & l_{o}
    \end{pmatrix}
\end{equation}

where $[d\,f\,g]_o/[d_o f_o g_o]$ is the cross product of $(h\,k\,l)_o$ and $[u\,v\,w]_o$ . The O value can also be derived by writing the rows as the cosine of the angle between the first $x^\#_c$ axis in Fig.~\ref{fig:f1} and the three sample axes {RD, TD, ND}. So on and so forth for $y_c$ and $z_c$.

A lot many times, there have been studies that centre around calculating the type of twinning in cubic to monoclinic stress-induced martensitic transformation in Ni-Ti shape memory alloys \citep{16, 17}. These studies compared the predictions by the Phenomenological Theory of Martensitic Crystallography (PTMC) to the experimentally calculated crystallographic parameters like habit plane, and orientation relationship for transformation in a single crystal Ni-Ti. Some used a concept of the minimum required rotation for the habit plane to be invariant in order to predict the best orientation relationship in $B2 \rightarrow B19'$ transformation \citep{18, 19}. A very sophisticated and generalised algorithm known as StrucTrans Algorithm was developed on the premise that stretch tensor U is the key variable that defines any structural transformation. Stretch tensor U is entirely dependent on the lattice parameter of the parent and daughter phase in the transformation and is used to develop a so-called Distance Function that predicts the feasibility of the lattice correspondence in the phase transformation \citep{11}. StrucTrans algorithm was written by the authors of this paper with input as the lattice parameter of cubic and monoclinic crystal system for phase transformation in Nickel-Titanium shape-memory alloy. The result gave different lattice correspondences with unique distance function values. Each lattice correspondence was in a set of 24 matrices which are related by the symmetry rotations in the cubic (c) point group $G^c$. For each lattice correspondence matrix in the set of 24 the value of Distance Function remains the same, as was also said in terms of Energy Wells \citep{12} that does not change for a stretch tensor U on a rigid body rotation or change of frame. The Distortion Matrix T is related to U by

\begin{equation}
    T^{T}T = U^2
\end{equation}

The T matrix also has 24 different values related by symmetry rotations in the cubic point group. But there are only 12 unique values of the stretch matrix that derives out of the Eq. 5. Amongst the 24 distortion matrices, 12 of them give the same U value as the remaining 12. Owing to the symmetry rotations in the point group of Monoclinic lattice $G^m$ the equation for the total number of orientational variants N by Lagrange's Formula becomes

\begin{equation}
N = \left | \frac{G^{c}}{G^{m}} \right |
\end{equation}

Cubic and Monoclinic crystal systems follow the group-subgroup relationship, and the number of orientational variants follows the Eq. 6 as pointed out by Janovec \citep{20, 21}. The fact that StrucTrans Algorithm gives 24 lattice correspondence is that it takes into account only the external symmetries of the parent phase into consideration. For different 24 distortion matrices T, 12 of them correspond to the same daughter crystal as the other 12 and are in coincidence. A very cogent argument on the number of orientational variants was made by Cyril Cyron \citep{22}.

The Distance function (D) \citep{11} is defined as

\begin{equation}
    D = \left \| (T^{T}T)^{-1} - I \right \|^{2}
\end{equation}

where $\left \| . \right \|$ represents the Frobenius norm. Different lattice correspondence with their respective D values for our particular lattice parameter is given in Table 1.

\begin{table}[h!]
\centering
\renewcommand{\arraystretch}{1.4}
\begin{tabular*}{ 0.5 \textwidth}{c c c c c}
\hline
Correspondence & $[1 0 0]$ & $[0 1 0]$ & $[0 0 1]$ & Distance \\
\hline
Involution & $[\frac{1}{2} \bar{\frac{1}{2}} \frac{1}{2}]$ & $[1 1 0]$ & $[\frac{1}{2} \bar{\frac{1}{2}} \frac{3}{2}]$  &  $0.0654$  \\
Bain & $[1 0 0]$ & $[0 1 1]$ & $[0 \bar{1} 1]$ & $0.0735$    \\
Other & $[\frac{1}{2} \frac{1}{2} \frac{1}{2}]$ & $[0 1 1]$ & $[\bar{1} 1 \bar{1}]$ & $0.1767$  \\
Other & $[\frac{1}{2} \frac{1}{2} \frac{1}{2}]$ & $[0 \bar{1} 1]$ &  $[\frac{3}{2} \bar{\frac{1}{2}} \bar{\frac{1}{2}}]$ & $0.2533$  \\
\hline
\end{tabular*}
\caption{Lattice Correspondence}
\label{table:t1}
\end{table}

As you will notice the correspondence with a D value equal to 0.0735 is the classic Bain Correspondence. But with a slightly less D value of 0.0654, we have another new lattice correspondence named as Involution Domain. Other lattice correspondences have a significantly large D value than the first two and are not feasible in a phase change from cubic to monoclinic. D value may be greater or less for Bain Correspondence in comparison to Involution Domain depending on the lattice parameter one chooses.

Twelve lattice correspondences are defined differently by Otsuka \citep{17} then the correspondence matrix $C^{\alpha\rightarrow\gamma}_{c}$ by Cyron \citep{14}. The correspondence matrix $C^{\alpha\rightarrow\gamma}_{c}$ changes any vector of parent (in parent reference basis) to a vector of daughter $\alpha$ (in daughter reference basis). So, it is a combination of two transformations; first, a distortion $F^{\gamma\rightarrow\gamma'}_{c}$ and second a coordinate transformation $T^{\alpha\rightarrow\gamma}_{c}$. Distortion matrix converts a vector in parent basis to some other distorted vector in the same basis. Then the distorted vector is written in the final daughter basis by doing a coordinate transformation. Again, both $F^{\gamma\rightarrow\gamma'}_{c}$ and $T^{\alpha\rightarrow\gamma}_{c}$ are derived in the same way as we derived S previously, but the application would be different following Eq. 20 and 21. Here the initial parent phase $\gamma$ first distorts to $\gamma’$ followed by a change of coordinate system to the final daughter phase $\alpha$. So, the symbol on F becomes $\gamma\rightarrow\gamma'$, while the symbol on T becomes $\alpha\rightarrow\gamma$ following the Eq. 21 $(T^{\alpha\rightarrow\gamma}_{c} = {(T^{\gamma\rightarrow\alpha}_{c})}^{-1})$.

\begin{equation}
    C^{\alpha\rightarrow\gamma}_{c} = T^{\alpha\rightarrow\gamma}_{c}F^{\gamma\rightarrow\gamma'}_{c}
\end{equation}

Correspondence matrix converts the vectors of parent crystal given in Table 1 to the three basis vectors of daughter phase (in daughter reference frame). The three vectors of parent crystal that convert to the basis vectors of the daughter crystal are called as Lattice Correspondence.

The distortion $F^{\gamma\rightarrow\gamma'}_{c}$ is defined by writing the three initial basis vectors after distortion as the columns of our matrix (reference frame remains the same). Though, $F^{\gamma\rightarrow\gamma'}_{c}$ is not a completely accurate representation for a general transformation. We have to take the actual lattice vectors, before and after the transformation (both in the same crystal reference frame). Cyron \citep{14} called this set of three lattice vectors as a supercell that distorts to the final set of lattice vectors. Kaushik Bhattacharya \citep{12} calculated exactly the same distortion matrix “T,” for cubic to monoclinic transformation, where \{$e^1_a$, $e^2_a$, $e^3_a$\} and \{$e^1_m$, $e^2_m$, $e^3_m$\} were both written in the cubic basis before and after the distortion of the lattice. Cyron otherwise defined this matrix to be in the “crystallographic basis” and defined it as

\begin{equation}
    F^{\gamma\rightarrow\gamma'}_{c,super} = B^{\gamma'}_{super}(B^{\gamma}_{super})^{-1}
\end{equation}

where $B^{\gamma}_{super}$ and $B^{\gamma'}_{super}$ are similar to \{$e^1_a$, $e^2_a$, $e^3_a$\} and \{$e^1_m$, $e^2_m$, $e^3_m$\} in \cite{12} respectively. Our Correspondence matrix then becomes

\begin{equation}
    C^{\alpha\rightarrow\gamma}_{c} = T^{\alpha\rightarrow\gamma}_{c}F^{\gamma\rightarrow\gamma'}_{c,super}
\end{equation}

The lattice correspondence for the Involution Domain given in Table 1 with D value of 0.0654, are the vectors that will change into the martensitic basis vectors after the distortion. In this report, we try to differentiate between the two domains based on the orientation data measured from EBSD. The method was initially made for Bain Domains, but we extended it to the Involution Domains after the authors of this report observed some anomalies in the results which pointed toward the possibility of Involution Domains.

\section{Experimental procedure}
\begin{figure}[h]
\begin{center}
    \includegraphics[scale = 0.3]{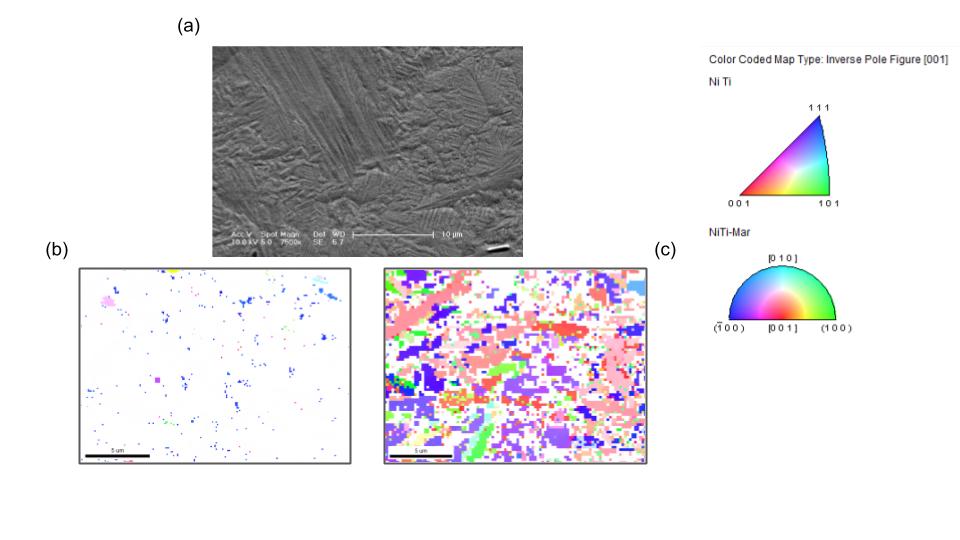}
    \centering
    \caption{EBSD map of annealed Ni-Ti-Cu sample.}
    \label{fig:f11}
\end{center}    
\end{figure}

\begin{figure}[h]
\begin{center}
    \includegraphics[scale = 0.4]{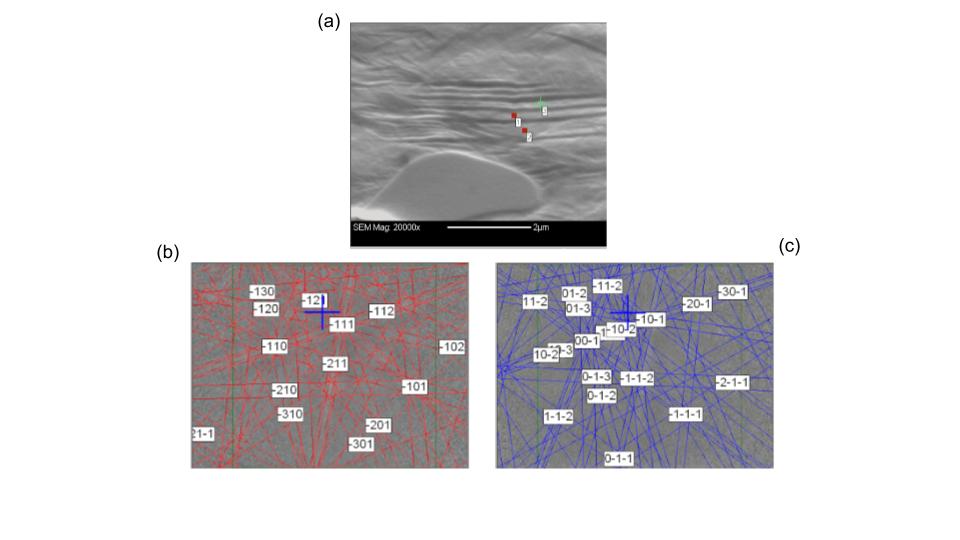}
    \centering
    \caption{(a) SE image (b) Kikuchi pattern for austenite (c) Kikuchi pattern for neighbouring martensite.}
    \label{fig:f12}
\end{center}    
\end{figure}

The Ni-Ti-Cu SMA having a nominal composition of Ni-47.2wt\% Ti -6.4wt\% Cu was prepared by vacuum arc melting technique by remelting them for at least 6 times for uniform composition. The alloy was solution treated at $1000^{\circ}$C for 24 hrs and then hot rolled at $900^{\circ}$C to 80\% thickness reduction. Further the rolled material was annealed at $600^{\circ}$C for 4 hours. The lattice parameter for the martensite phase was obtained as a=0.291 nm, b=0.41306 nm, c=0.46818 nm and $\beta=98.09^{\circ}$ using X-ray diffraction patterns obtained from Rigaku Smart Lab Diffractometer. The lattice parameter for the austenite phase were taken as 0.3013 nm as obtained from literature for similar composition. Sample for electron back scattered diffraction (EBSD) was prepared from the rolled material following standard metallographic procedure and electropolishing. The EBSD was acquired using Oxford Nordlys Nano detector attached to a Carl Zeiss Gemini 300 Field Electron Gun Scanning Electron microscope at 25 kV and 2 nA with a step size of 40 nm. Data acquistion was carried out at $80^{\circ}$C using Kammrath Weiss in-situ heating holder to have combined austenite and martensite phases. The transformation temperatures of the sample obtained from differential scanning calorimetry are 23, 52, 60, 92 $^{\circ}$C  respectively. Fig.~\ref{fig:f11}a shows the secondary electron image of the annealed sample depicting the twinned martensitic structure. The orientation maps of the austenite and martensite regions are given in Fig.~\ref{fig:f11}b and (c) respectively. The Kikuchi patterns of the austenite and neighbouring martensite regions are shown in Fig.~\ref{fig:f12}. The maximum angular deviation for both of these patterns are less than $1^{\circ}$ indicating a reasonable indexing of the patterns. Since many of the martensite twins are having dimensions smaller than 40 nm, i.e. step size of the EBSD scan, it was not possible to index those twins. 

\section{Methodology}

\begin{figure}[h]
\begin{center}
    \includegraphics[scale = 0.1]{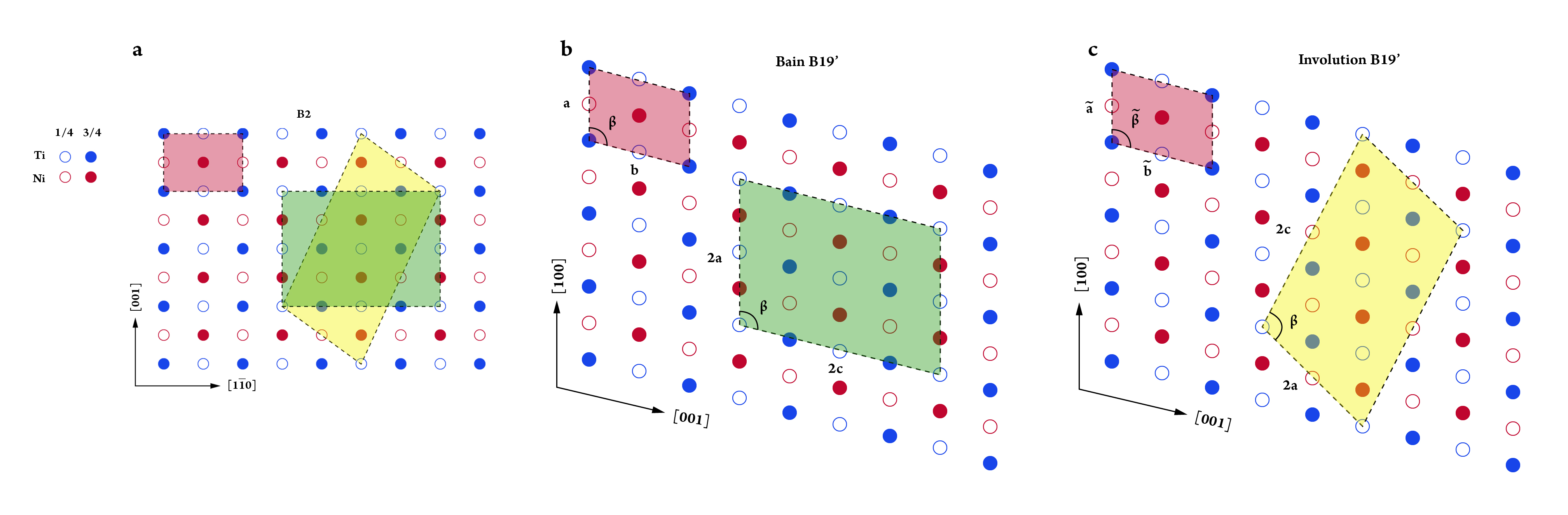}
    \centering
    \caption{(a) The $(\bar{1} \bar{1} 0)$ projection of B2 lattice where the green and yellow super cell converts to (b) $(0 \bar{1} 0)$ B19' Bain projection and (c) $(0 1 0)$ B19' Involution projection, respectively.}
    \label{fig:f2}
\end{center}    
\end{figure}

To better understand both, the Bain and Involution Domains, Fig.~\ref{fig:f2} \citep{10} shows the lattice correspondence in the cubic reference frame that converts to the basis vector of monoclinic crystal. It shows one of the 12 variants for both Bain and Involution Domains. When we talk about “variants,” it is a relative terminology. A variant of martensite depends on both the orientation of austenite and martensite. Let us assume for a moment that in our Fig.~\ref{fig:f2}a is a crystal of Nickel-Titanium in real space. Now,  try to imagine its orientation with Normal Direction (ND) pointing vertically upwards in the plane of the paper while Rolling Direction (RD) points horizontally out of the plane. For this particular variant, the orientation matrix for austenite must be 

\begin{equation}
    O_{a}=\begin{pmatrix}
    -1/\sqrt{2} & 1/\sqrt{2} & 0\\ 
    -1/\sqrt{2} & -1/\sqrt{2} & 0\\ 
    0 & 0 & 1
    \end{pmatrix}
\end{equation} 

In accordance with one of the lattice correspondence, the martensitic basis for both Bain and Involution Domain is given in Fig.~\ref{fig:f2}b and Fig.~\ref{fig:f2}c respectively. The orientation matrix $O_m$ for the corresponding martensite domains can be calculated once we know the correct lattice correspondence or vice versa. We can calculate the lattice correspondence between austenite and martensite once we have the orientation of both the phases. The orientation matrices can be connected to each other by a simple coordinate transformation as

\begin{equation}
    O_a = GO_m
\end{equation}

G is a transformation matrix derived by writing the orthogonal reference connected to the monoclinic basis, in the initial cubic reference frame. The G matrix can be calculated in three simple steps:-

\begin{enumerate}
    \item Using Eq. 1 and Eq. 2 write the orthogonal reference basis $\{x^\#_c, y^\#_c, z^\#_c\}$ into the monoclinic reference frame.
    \item Change the reference from monoclinic to the cubic frame using $T^{\gamma\rightarrow\alpha}_{c}$
    \item Normalize the columns.
\end{enumerate}

\begin{equation}
    G_{bain} = T^{\gamma\rightarrow\alpha}_{c}{A}^{-1}
\end{equation}

\begin{center}
$G_{bain} = \begin{pmatrix}
0 & -b/\sqrt{2} & c\sin \beta /\sqrt{2}\\ 
0 & -b/\sqrt{2} & -c\sin \beta /\sqrt{2}\\ 
a & 0 & c\cos \beta 
\end{pmatrix}
\begin{pmatrix}
a\sin\beta & 0 & 0\\ 
0 & b & 0\\ 
a\cos\beta  & 0 & c
\end{pmatrix}^{-1}$
\end{center}

where a, b, c and $\beta$ are the lattice parameters for the monoclinic crystal. On substituting the lattice parameters for Ni-Ti-Cu we obtained from the experimental procedure

\begin{equation}
    G_{bain} = \begin{pmatrix}
    0.0995 & -0.7071 & 0.7001\\ 
    0.0995 & -0.7071 & -0.7001\\ 
    0.9900 & 0 & -0.1407 
    \end{pmatrix}  
\end{equation}

The transformation matrix $G_{bain}$ derived above is for only one variant of Bain Domain as shown in Fig.~\ref{fig:f2}b. Similarly, there will be a total of 24 symmetry related $G_{bain}$ matrices. Twenty four classic Bain variants have been listed in Table 2. If the sample is now given a random rotation R in {RD, TD, ND}  so as the orientation $O_a$ and $O_m$ also rotate by R the Eq. 12 modifies to 

\begin{equation}
    RO_a = RGO_m
\end{equation}

Still, the transformation matrix G remains the same, therefore pointing at the same variant. $T^{\gamma\rightarrow\alpha}_{c}$ value is different for different lattice correspondence in Table 2. For any orientation of austenite and martensite phase on applying Eq. 12 we will get one of the G values calculated for Bain Domain. For example, as we suggested earlier that the variants are relative, we could have numbered the lattice correspondence in some other sequence. In the case of twins, the variants of martensite could be 1-10 or 5-6 if the same lattice correspondence were to be numbered 5 and 6 instead of 1 and 10. But at last, the twin plane will come out to be the same for both combinations, because the stretch matrix U will remain exactly the same. Swapping of adjacent lattice correspondences in Table 2 won’t make any difference in the actual variant. So, 1 can be swapped with 13, or 5 can be swapped with 17 because both 1,13 and 5,17 have the same U values and therefore correspond to the same variant.

\begin{table}[h]
\centering
\renewcommand{\arraystretch}{1.2}
\begin{tabular}{ll}
\hline
\begin{tabular}{c c c c}
Variant & $[1 0 0]_m$ & $[0 1 0]_m$ & $[0 0 1]_m$ \\
\cline{1-4}
1 & $[1 0 0]_p$ & $[0 \bar{1} {1}]_p$ & $[0 \bar{1} \bar{1}]_p$ \\
2 & $[\bar{1} 0 0]_p$ & $[0 {1} \bar{1}]_p$ & $[0 \bar{1} \bar{1}]_p$ \\
3 & $[1 0 0]_p$ & $[0 {1} \bar{1}]_p$ & $[0 {1} {1}]_p$ \\
4 & $[\bar{1} 0 0]_p$ & $[0 \bar{1} {1}]_p$ & $[0 {1} {1}]_p$ \\
5 & $[0 1 0]_p$ & $[{1} 0 {1}]_p$ & $[{1} 0 \bar{1}]_p$ \\
6 & $[0 \bar{1} 0]_p$ & $[\bar{1} 0 \bar{1}]_p$ & $[{1} 0 \bar{1}]_p$ \\
7 & $[0 1 0]_p$ & $[\bar{1} 0 \bar{1}]_p$ & $[\bar{1} 0 {1}]_p$ \\
8 & $[0 \bar{1} 0]_p$ & $[{1} 0 {1}]_p$ & $[\bar{1} 0 {1}]_p$ \\
9 & $[0 0 1]_p$ & $[\bar{1} \bar{1} 0]_p$ & $[{1} \bar{1} 0]_p$ \\
10 & $[0 0 \bar{1}]_p$ & $[{1} {1} 0]_p$ & $[{1} \bar{1} 0]_p$ \\
11 & $[0 0 1]_p$ & $[{1} {1} 0]_p$ & $[\bar{1} {1} 0]_p$ \\
12 & $[0 0 \bar{1}]_p$ & $[\bar{1} \bar{1} 0]_p$ & $[\bar{1} {1} 0]_p$ \\
\cline{1-4}
\end{tabular}
&
\begin{tabular}{c c c c}
\renewcommand{\arraystretch}{1.2}
Variant & $[1 0 0]_m$ & $[0 1 0]_m$ & $[0 0 1]_m$ \\
\cline{1-4}
$13\leftrightarrow1$ & $[\bar{1} 0 0]_p$ & $[0 \bar{1} {1}]_p$ & $[0 {1} {1}]_p$ \\
$14\leftrightarrow2$ & $[{1} 0 0]_p$ & $[0 {1} \bar{1}]_p$ & $[0 {1} {1}]_p$ \\
$15\leftrightarrow3$ & $[\bar{1} 0 0]_p$ & $[0 {1} \bar{1}]_p$ & $[0 \bar{1} \bar{1}]_p$ \\
$16\leftrightarrow4$ & $[{1} 0 0]_p$ & $[0 \bar{1} {1}]_p$ & $[0 \bar{1} \bar{1}]_p$ \\
$17\leftrightarrow5$ & $[0 \bar{1} 0]_p$ & $[{1} 0 {1}]_p$ & $[\bar{1} 0 {1}]_p$ \\
$18\leftrightarrow6$ & $[0 {1} 0]_p$ & $[\bar{1} 0 \bar{1}]_p$ & $[\bar{1} 0 {1}]_p$ \\
$19\leftrightarrow7$ & $[0 \bar{1} 0]_p$ & $[\bar{1} 0 \bar{1}]_p$ & $[{1} 0 \bar{1}]_p$ \\
$20\leftrightarrow8$ & $[0 {1} 0]_p$ & $[{1} 0 {1}]_p$ & $[{1} 0 \bar{1}]_p$ \\
$21\leftrightarrow9$ & $[0 0 \bar{1}]_p$ & $[\bar{1} \bar{1} 0]_p$ & $[\bar{1} {1} 0]_p$ \\
$22\leftrightarrow10$ & $[0 0 {1}]_p$ & $[{1} {1} 0]_p$ & $[\bar{1} {1} 0]_p$ \\
$23\leftrightarrow11$ & $[0 0 \bar{1}]_p$ & $[{1} {1} 0]_p$ & $[{1} \bar{1} 0]_p$ \\
$24\leftrightarrow12$ & $[0 0 {1}]_p$ & $[\bar{1} \bar{1} 0]_p$ & $[{1} \bar{1} 0]_p$ \\
\cline{1-4}
\end{tabular}
\end{tabular}
\caption{Bain Correspondence}
\label{table:t2}
\end{table}

The transformation matrix $G_{bain}$ will correspond to Bain Domain. $G_{invo}$ can also be calculated based on the lattice correspondence for the Involution Domain. Calculation of $G_{invo}$ is not very straight forward as we did in Eq. 13. Although, the Structure Tensor A will remain the same, it is not possible to calculate $T^{\gamma\rightarrow\alpha}_{c}$ for Involution directly as the angles between the martensitic basis in Fig.~\ref{fig:f2}c and the cubic basis are not know. With a little help of matrix and vector algebra, the calculation of $G_{invo}$ can be done. The $G_{invo}$ matrix for the particular variant in Fig.~\ref{fig:f2}c and the same lattice parameters used for Bain Domain comes out as:

\begin{equation}
    G_{invo} = \begin{pmatrix}
    0.6153 & 0.7071 & 0.3484\\ 
    -0.6153 & 0.7071 & -0.3484\\ 
    -0.4927 & 0 & 0.8702 
    \end{pmatrix}  
\end{equation}

The modulus of the indices (keeping the lattice parameters unchanged) depends on the Structure Tensor A used. Change in A may also result in the interchanging of the columns of the G  matrix, and one may get different answers out Eq. 12.

\section{Result and Discussion}

EBSD data from a Ni-Ti-Cu sample was studied using TSL OIM Analysis Software. Orientation matrix calculated from multiple sites on the IPF map was exported. The orientation was calculated for Austenite and Martensite were adjacent or close to each other. The orientation matrices for different sites are given below. For each site, we checked by applying Eq. 12 to understand the type of martensite present.

Site 1:

\begin{center}
$O_{a} = \begin{pmatrix}
0.540 & -0.648 & 0.537\\ 
0.316 & 0.747 & 0.584\\ 
-0.780 & -0.145 & 0.608
\end{pmatrix}\; \; \; \; O_{m} = \begin{pmatrix}
0.834 & 0.167 & -0.526\\ 
-0.150 & 0.986 & 0.075\\ 
0.531 & 0.016 & 0.847
\end{pmatrix}$ 
\end{center}

\begin{center}
    $G = \begin{pmatrix}
    0.0597 & -0.6792 & 0.7312\\ 
    0.0810 & 0.7329 & 0.6749\\ 
    -0.9944 & 0.0198 & 0.0985
    \end{pmatrix}$
\end{center}

The G matrix value matches quite closely to the one of the 24 $G_{bain}$ matrices showing that the martensite present is in Bain Domain.

Site 2:

\begin{center}
$O_{a} = \begin{pmatrix}
0.872 & -0.468 & 0.139\\ 
0.463 & 0.884 & 0.072\\ 
-0.157 & 0.002 & 0.988
\end{pmatrix}\; \; \; \; O_{m} = \begin{pmatrix}
0.839 & 0.271 & -0.472\\ 
-0.544 & 0.420 & -0.726\\ 
0.002 & 0.866 & 0.500
\end{pmatrix}$ 
\end{center}

\begin{center}
    $G = \begin{pmatrix}
    0.5393 & -0.7725 & -0.3345\\ 
    0.5937 & 0.0674 & 0.8023\\ 
    -0.5976 & -0.6312 & 0.4954
    \end{pmatrix}$
\end{center}
On observing the columns it can be concluded that it is one of the variants of Involution Domain. The  G matrix value is significantly close to one of the 24 transformation matrices $G_{invo}$.

\begin{table}[h]
\centering
\renewcommand{\arraystretch}{1.2}
\begin{tabular}{>{\centering\arraybackslash}m{0.8cm} >{\centering\arraybackslash}m{2.9cm} >{\centering\arraybackslash}m{2.9cm} >{\centering\arraybackslash}m{2cm} >{\centering\arraybackslash}m{2cm}}
\hline
Site & Autenite Orientation & Martensite Orientation & Variant Type & Misorientation Angle \\
\hline
1 & (1.618  1.761  1.833) [0.179  0.105  $\overline{0.259}$] & ($\overline{1.862}$  0.311  3.967) [0.290  $\overline{0.036}$  0.139] & Bain Type & 2.87° \\
2 & (0.420  0.217  2.976) [0.290  0.154  $\overline{0.052}$] & ($\overline{1.565}$  $\overline{2.998}$  2.343) [0.291  $\overline{0.132}$  0.026] & Involution Type & 8.6° \\
3 & (0.541  0.886  2.829) [0.280  $\overline{0.178}$  0.002] & ($\overline{1.688}$  3.253  $\overline{0.584}$) [0.268  0.149  0.056] & Involution Type & 7.77° \\
4 & (1.779  1.555  1.870) [$\overline{0.262}$  0.176  0.103] & (1.754  $\overline{0.415}$  3.272) [$\overline{0.237}$  0.038  0.132] & Bain Type & 2.11° \\
5 & (1.788  0.429  2.387) [$\overline{0.253}$  0.139  0.164] & (1.587  $\overline{2.970}$  $\overline{2.358}$) [$\overline{0.284}$  $\overline{0.139}$  $\overline{0.016}$] & Bain Type & 5.45° \\
6 & (1.788  0.429  2.387) [$\overline{0.253}$  0.139  0.164] & (0.433  $\overline{0.828}$  $\overline{4.587}$) [0.153  $\overline{0.213}$  0.053] & Bain Type & 2.90° \\
\hline
\end{tabular}
\caption{Miller Indices of Austenite and close Martensitic sites in Nickel-Titanium-Copper sample.}
\label{table:t3}
\end{table}

\begin{figure}[h]
\begin{center}
    \includegraphics[scale = 0.4]{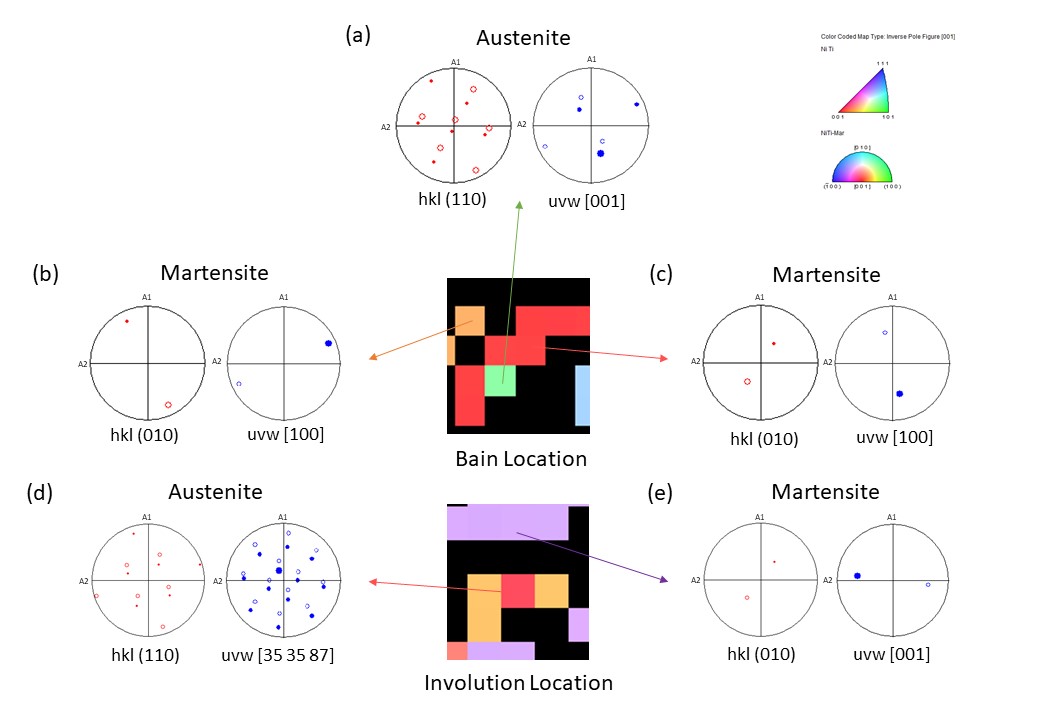}
    \centering
    \caption{IPF map of Bain location corresponding to 5th and 6th site in Table 3, pointing to it's respective austenite (a) and martensite (b), (c) pole figures. IPF map of Involution location corresponding to 2nd site in Table 3 with (d) and (e) as it's respective austenite and martensite pole figures. }
    \label{fig:f3}
\end{center}    
\end{figure}

More results inclusive of the above Site 1 and Site 2 have been shown in Table 4. The orientation in Table 4 is given in the form of Miller Indices which can be converted to the Orientation matrix using the Eq. 2 and 3. The pole figures for some of these sites in Table 4 are shown in the Fig.~\ref{fig:f3}. 

$G_{invo}$ matrix was calculated purely through vector algebra with the help of Fig.~\ref{fig:f2}c. But, there was an important observation that needs to be pointed out. In the next few lines, we will explain the calculation in a summary. For the calculation to be more incisive we took an Austenite orientation to be (1 1 0)[0 0 1]. Rotating it by 60 degrees about [-1 1 1] gave us a new austenitic orientation as (1 0 1)[-2 -1 2]. Subsequently, by applying the Eq 12, for all possible symmetry-related $G_{bain}$ matrices we calculated 24 possible orientations for martensite. When calculated the G matrices using the final martensite orientations and the initial austenite orientation (1 1 0)[0 0 1] we got many G matrices close to $G_{invo}$ one of which is as

\begin{equation}
    G\:matrix = \begin{pmatrix}
    -0.6269 & -0.7071 & 0.3272\\ 
    -0.4627 & 0 & -0.8865\\ 
    0.6269 & -0.7071 & -0.3272 
    \end{pmatrix}  
\end{equation}

The above matrix is nothing but yet another symmetry related variant of $G_{invo}$. A slight change in the indices of the above G matrix to the $G_{invo}$ is caused by the error in trigonometric functions that we used for calculation of $G_{invo}$.

The conclusion that we made from the above output points us in a direction where the austenite undergoes a twinning before it converts into martensite. Transformation to Involution type can be understood as a two step process here.

\subsection{Perturbation in the Data}

After calculating the G matrix value at a copious number of sites in Ni-Cu-Ti samples and also some Ni-Ti samples, we observed a consistent rotation in the G matrix value of the involution domain. The  $G_{invo}$ was systematically rotated by approximately 8°- 9° anti-clockwise about the $y_c$ axis of the monoclinic crystal (pointing inside the plane). The G matrix calculated experimentally at two of the sites with austenite and martensite phase close to each other in Ni-Ti alloy are given below:

Site 1:

\begin{center}
    $G = \begin{pmatrix}
    -0.6661 & -0.7052 & -0.2436\\ 
    0.6567 & -0.7103 & 0.2543\\ 
    -0.3524 & 0.0094 & 0.9366
    \end{pmatrix}$
\end{center}

Site 2:

\begin{center}
    $G = \begin{pmatrix}
    -0.6431 & -0.7070 & -0.2926\\ 
    0.6606 & -0.7060 & 0.2548\\ 
    -0.3854 & -0.0291 & 0.9220
    \end{pmatrix}$
\end{center}

A plethora of sites in Ni-Ti alloy showed this kind of perturbation and also some sites in Ni-Ti-Cu. The anomaly is very consistent all over the sample and also with different alloys. It cannot be explained as only an experimental error in the orientation matrix from the EBSD pattern. However, we noted an interesting point. There has been this theory based on DFT(Density Functional Theory) calculation where the angle associated with minimum energy distortion is ~107° rather than ~98.5° what we experimentally observe \citep{23}. It can be noted in Fig.~\ref{fig:f2}c the angle between the martensite $[1 0 0]_{B19'}$ basis and martensite  $[0 0 1]_{B19'}$ basis is around 98.5°. The structure matrix A in our software (TSL OIM Analysis) is defined by keeping the c-axis of monoclinic parallel to the z-axis of the attached cartesian system, in order to calculate the orthonormal orientation matrix from the Miller Indices. A conjecture that can be drawn for the rotation of the G matrix is that monoclinic c-axis in the cubic reference frame, i.e. [0.3484 -0.3484 0.8702] should be rotated by 8°- 9° anti-clockwise about the monoclinic b-axis (pointing inside the plane). Due to this rotation, the angle $\beta$ becomes approximately 107°. The misorientation between the $G_{invo}$ and perturbed G was calculated using the following equation

\begin{equation}
    \delta G = G_{invo}(G)^{-1} 
\end{equation}

\begin{equation}
    Misorientation Angle = \arccos\left[\frac{(trace(\delta G)-1)}{2}\right]
\end{equation}

\section{Conclusion}

We have a theory that can be used to identify the type and variant of martensite quickly from the EBSD data. Lattice correspondence has a connection with the orientation matrices of different phases and thus can be useful in predicting the type of martensite variant. Once the type and variant are known subsequently, we could calculate the twin planes and habit planes to understand more about the compatibility between different phases. We will be discussing these in our upcoming publication. Also, further investigation is needed for the $G_{invo}$ in order to identify the Involution Domains much more accurately.

Acknowledgement: The authors greatly acknowledge the funding provided by the Department of Science and Technology, Government of India (Grant No: DST/SERB/ECR/2016/000883)

\bibliographystyle{abbrvnat}

\appendix

\section{Appendix}
For a displacive transformation, defining lattice vectors and transformation matrices is a crucial thing. We will be defining some of the transformation matrices like coordinate transformation, vector transformation matrices (also known as distortion matrices) \cite{13,14}. A crystal system is defined by six independent constants a, b, c, $\alpha$, $\beta$, $\gamma$ called as the Lattice Parameter. The a, b, c are the length of the three basis vectors and $\alpha$, $\beta$, $\gamma$ the angles between them. The $\alpha$ is angle between (b and c), $\beta$ is the angle between (c and a), and $\gamma$ is the angle between (a and b).

\begin{figure}[h]
\begin{center}
    \includegraphics[scale = 0.5]{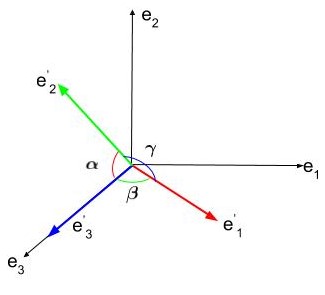}
    \centering
    \caption{Colored $\{e_1', e_2', e_3'\}$ represents a general coordinate system while black $\{e_1, e_2, e_3\}$ represents an orthonormal coordinate system.}
    \label{fig:f4}
\end{center}    
\end{figure}

Let us consider the two crystal systems with crystallographic basis represented by {$e_1$, $e_2$, $e_3$} and {$e_1'$, $e_2'$, $e_3'$} with the former as an initial orthonormal system and later as a final general crystal system. The latter has angles $\alpha$, $\beta$ and $\gamma$  between their basis vectors with $|e_1'|$  = a, $|e_2'|$ = b, $|e_3'|$ = c as shown in Fig.~\ref{fig:f4}.

The transformation matrix $S_i^f$ can be defined by expressing the final (f) basis $\{e_1', e_2', e_3'\}$ in the initial (i) orthogonal basis $\{e_1, e_2, e_3\}$ . With a little geometry and vector calculations \cite{13} the columns of the matrix $S_i^f$ comes out to be as \\* \\*
$s_1^{1} = a\sin\beta$ \\*
$s_1^{2} = 0$ \\*
$s_1^{3} = a\cos\beta$ \\*
$s_2^{1} = \frac{b}{\sin\beta}(\cos\gamma - \cos\alpha\cos\beta)$ \\*
$s_2^{2} = \frac{b}{\sin\beta}(2\cos\gamma\cos\alpha\cos\beta+\sin^{2}\alpha sin^{2}\beta - cos^{2}\alpha cos^{2}\beta - cos^{2}\gamma)$ \\*
$s_2^{3} = b\cos\alpha$ \\*
$s_3^{1} = 0$ \\*
$s_3^{2} = 0$ \\*
$s_3^{3} = c$ \\*

where

\begin{center}
\begin{math}
S_{f}^{i} = \begin{pmatrix}
s_1^{1} & s_2^{1} & s_3^{1}\\ 
s_1^{2} & s_2^{2} & s_3^{2}\\ 
s_1^{3} & s_2^{3} & s_3^{3}
\end{pmatrix}
\end{math}
\end{center}

Because one of the crystal bases, $\{e_1', e_2', e_3'\}$, is not an orthogonal system, we have to be very careful in defining the $S_i^f$ matrix. $S_i^f$ is not an orthogonal matrix in our case and therefore $(S_i^f)^T \neq (S_i^f)^{-1}$.

Any vector x in the initial frame of reference which we defined by orthonoral basis $\{e_1, e_2, e_3\}$ distorts to a new vector x’ in the same reference frame, by 

\begin{equation}
   x' =  S_i^fx   
\end{equation}

While any vector u can be defined in two different coordinate systems by

\begin{equation}
   u =  S_i^fu'   
\end{equation}

where u is in the $\{e_1, e_2, e_3\}$ crystal bases and u’ is in $\{e_1', e_2', e_3'\}$ crystal bases. Note, the vector u does not change in the real space.

\end{document}